# Comments on chemical bonding in the high pressure form of Boron

Piero Macchi,

Department of Chemistry and biochemistry,

University of Bern, Freiestrasse 3, 3012 Bern Switzerland

Recently, the high pressure structure of Boron ($\gamma$-B) has been structurally characterized by Oganov and coworkers[1] and by Zarechnaya and coworkers.[2] Although the structure reported by the two research groups is basically the same, some controversy was raised especially concerning the interpretation of the chemical bonding in this new phase.

Oganov *et al.* determined the structure of $\gamma$-B using an *evolutionary algorithm*, and confirmed the structure by X-ray diffraction. The interesting feature is the presence of $B_2$ "dumbbells" connected with the more common $B_{12}$ icosahedra. Upon quantum mechanical calculations in the solid state, those authors concluded that there is a partial charge transfer from $B_2$ to $B_{12}$, and they called the compound a "boron-boride". As there are several possible atomic partitioning of the electron density, the charge transfer is not uniquely determined, but Oganov *et al.*[1] reported qualitatively similar conclusions from different definitions of atomic charge. A *ionic* structure would be quite interesting, because showing that the pressure is able to stabilize electronic configurations (and corresponding structures) that are not stable at room conditions. Because the field of high pressure structural chemistry is relatively young, it can be envisaged that much theoretical research will be dedicated in the next few years to explain new chemical bonding features. Oganov *et al.*[1] have stressed also on the ionic component of one of the bonds between the two subunits.

Zarechnaya *et al.*[2] reported single crystal and powder X-ray diffraction data, together with theoretical calculations on $\gamma$-Boron. Beside the structural features are substantially similar to those of Oganov *et al.*,[1] the controversy is in part due to the different interpretation of the chemical bond between $B_{12}$ and $B_2$ units. After reading the two papers, it seems necessary to discuss some terminology and to address possible sources of confusion, which is in part due to different languages adopted by chemists and physicists. Zarechnaya *et al.*,[2] for example, seems to automatically assign the nature of the bond between two atoms (or molecular groups) based on the electronic configuration only. In this view, a bond between two *ions* would be necessarily *ionic*. This is, however, a dangerous oversimplification.

A glimpse at some well-known compound would immediately clarify the problem. We can focus, for example, on the coordination compounds involving transition metal atoms and organic or inorganic (closed shell) molecules or ions. Apart for the sub-class of zerovalent metals, all these compounds are formed by metal cations and neutral or anionic ligands. Applying the oversimplification described above, one shall conclude that all these compounds are characterized by ionic metal-ligand bonds, which was in fact the initial interpretation that prompted to the so-called *crystal field theory* (CFT). With the CFT, the spectroscopic properties of transition metals is explained by the "d-orbital splitting" occurring to an ion surrounded by negative charges (breaking the spherical symmetry of the ion). No other interaction is assumed in this model. Since more than 50 years, however, we know that this is not correct and that (substantial) covalent bonding has to be considered when describing these species, unless missing some important information (one of the first evidence being the *nephelauxetic effect*). Thanks especially to the work by B. N. Figgis,[3] a *ligand field theory* (LFT) was later developed, still valid within the molecular orbital frameworks, extensively used afterwards for more accurate calculation of the electronic states of these species.

The changes observed for Boron at high pressure are not unique. An interesting example of a organic compound is oxalic acid di-hydrate (see Casati *et al.*[4]), which at high pressure transforms into oxalate di-anion and hydronium species (because the electronic configuration is mainly assigned based on the location of an Hydrogen atom, the formation of ionic units is much less ambiguous here). Again, the oversimplification of Zarechnaya *et al.*[2] would lead to think that the

(hydrogen) bond between the two units is exclusively ionic at high pressure. On the other hand, the hydrogen bond is an electrostatic interaction when it is weaker (for example at room pressure, where neutral oxalic acid and water molecules are more weakly interacting) but it becomes more covalent as the donor and acceptor are (forced to) approaching (for example, in the high pressure form). Prof. Gilli and co-workers have brilliantly illustrated which are the conditions that more likely give covalent hydrogen bonding (hence shorter donor-acceptor distances).[5] The covalency of a X-H---Y system is always due to the interplay between the two covalent forms (a neutral and a charge-transfer one) associated with the valence bond description of a hydrogen bonding, namely X-H---Y and (-)X---H-Y(+). The more these configurations interact (meaning the closer they are in energy), the higher is the covalency.

Given this illustration, it might become clear that finding features of covalent bonding between the $B_2$-$B_{12}$ units in γ-B is not in contradiction with a ionic electronic configurations. In a more recent report (http://arxiv.org/abs/0907.1900) Zarechnaya et al.[2] stress that there is a discontinuity between polar covalent bonds and ionic bonds. This is not so: ionic bonds are simply an extreme case of a polar bond. In addition, it shall be stressed that covalency and ionicity are not mutually exclusive. This is well known in chemistry. In recent theoretical work by the group of prof. Martin Pendas, an interpretation scheme was developed based on the electron distribution function (a probabilistic approach). They showed how some bonds can be at the same time strongly covalent and strongly ionic. A notable example being carbon monoxide (CO), a molecule with a formal triple bond (hence strongly covalent) but also with a strong bond polarity due to the different electronegativity of C and O (note that the small dipole moment of this species is not in contradiction with the large bond dipole, in fact, the molecular dipole is due to the large non bonding charge concentration of the inert lone pair on the C atom which almost oppose the large bond dipole).

Calculations reported by Zarechnaya et al.[2] are not necessarily dismissing the electronic configuration of $B_{12}$-$B_2$ units reported by Oganov and coworkers.[1] Zarechnaya et al.[2] do not report results of charge partitioning, so we do not know whether their theoretical approach is in fact speaking of a different electron density distribution or not. Oganov et al.[1] always assumed the basic covalent skeleton in γ-B, and simply addressed the larger ionic character in one specific B-B bonds. This was clarified more recently in an addendum in *Nature*, by the same authors.

The definition of covalency and the deformation density approach used by Zarechnaya et al.[2] might require some careful attention. Covalency is not the presence of "interstitial charge density" between two atoms, rather covalency is associated with a substantial *electron sharing* between two atoms. While electron distribution is easily measurable, electron sharing is not. Indeed, it might produce charge accumulation in a bond, but this is not always the case.[6] Moreover, it is quite well known in the community of charge density analysis that the difference density between a total and a reference electron density (often taken as the sum of isolated atoms) might lead to ambiguities and misinterpretations. Well known since the early '80s are the examples of $F_2$ and $H_2O_2$, molecules that would appear non-covalent, because lacking of charge accumulation in the F-F or O-O bonds. As it was proven quite clearly, this lack is just an artifact due to the non-spherical ground state of the isolated F or O atoms. The choice to use the sphericized atomic densities as reference leads to charge lacking in the bonding regions. Notably, a similar problem (but of opposite sign) would affect Boron. Neither Oganov et al.[1] nor Zarechnaya et al.[2] specify what are the electronic configuration and functions used to compute the independent atom model density used as reference in the deformation maps or profile reported in their corresponding papers, so we do not know exactly how much this problem might affect the maps. Very likely, the sphericized $^2P_{1/2}$ ground state electron density of B in $1s^22s^22p^1$ configuration was used.

Instead of the deformation density approach, in the past 20 years (or more) the charge density community has much privileged the analysis of the total electron density, within the framework of the quantum theory of atoms in molecules (QTAIM) as developed by Bader.[7] Curiously, in their online report Zarechnaya et al.[2] called the deformation density approach a "Bader analysis", which is certainly improper, as prof. R. Bader has always (and vigorously) dismissed the deformation density method. Oganov *et al.*[1] analyzed the total electron density and showed a clear asymmetry (due to bond polarity) in the bond between the two subunits, although further analysis would be necessary to better ascertain the nature of this bond. The ELF approach used by Zarechnaya *et al.*[2] points out that there is substantial electron localization in B-B bond. Importantly, these two pieces of information are not in contradiction with each other.

In conclusion, some of the observations reported by two groups of scientists and related to the atomic and electronic structure of the high pressure form of Boron do not seem to be contradiction to each other, although two different perspectives were used. On the other hand, some clarification of the terminology used is necessary sometime to better characterize the bonding properties of materials.

**Acknowledgements**
The Swiss National Science Foundation is acknowledged (project 200021_126788/1)

**References**

[1] A. R. Oganov, J. Chen, C. Gatti, Y. Ma, Y. Ma, C. W. Glass, Z. Liu, T. Yu, O. O. Kurakevych, V.r L. Solozhenko *Nature* **2009**, *457*, 863.
[2] E.Yu. Zarechnaya, L. Dubrovinsky, N. Dubrovinskaia, Y. Filinchuk, D. Chernyshov, V. Dmitriev, N. Miyajima, A. El Goresy, H. F. Braun, S. Van Smaalen, I. Kantor, A. Kantor, V. Prakapenka, M. Hanfland, A. S. Mikhaylushkin, I. A. Abrikosov, S. I. Simak *Phys. Rev. Lett.* **2009**, *102*, 185501.
[3] B. N. Figgis *Ligand Field Theory and Its Applications*, Wiley-VCH, **2000**
[4] N. Casati, P. Macchi, A. Sironi *Chem. Commun.* **2009**, 2679
[5] G. Gilli, P. Gilli, *Journal of Molecular Structure*, **2000**, *552*, 1
[6] P. Macchi, A. Sironi, *Coord. Chem. Rev.* **2003**, *238-239*, 383.
[7] Bader, R. F. W. *Atoms in Molecules* Oxford University Press: Oxford, U.K., **1990**.